\documentclass[sigconf, authorversion, nonacm]{acmart}
\usepackage{caption}
\usepackage{subcaption}
\usepackage{balance}
\usepackage{lipsum}%
\PassOptionsToPackage{showframe}{geometry}
\usepackage[showframe]{geometry}
\usepackage{array, caption, tabularx,  ragged2e,  booktabs}

\AtBeginDocument{%
  \providecommand\BibTeX{{%
    \normalfont B\kern-0.5em{\scshape i\kern-0.25em b}\kern-0.8em\TeX}}}

\begin{document}

\title{Evaluating Deception and Moving Target Defense with Network Attack Simulation}

\author{Daniel Reti}
\affiliation{%
  \institution{German Research Center for Artificial Intelligence}
  \city{Kaiserslautern}
  \country{Germany}}
\email{daniel.reti@dfki.de}

\author{Karina Elzer}
\affiliation{%
  \institution{German Research Center for Artificial Intelligence}
  \city{Kaiserslautern}
  \country{Germany}}
\email{karina.elzer@dfki.de}

\author{Daniel Fraunholz}
\affiliation{%
  \institution{Independent Researcher}
  \city{Munich}
  \country{Germany}}
\email{daniel.fraunholz@gmx.de}

\author{Daniel Schneider}
\affiliation{%
  \institution{German Research Center for Artificial Intelligence}
  \city{Kaiserslautern}
  \country{Germany}}
\email{daniel.schneider@dfki.de}

\author{Hans Dieter Schotten}
\affiliation{%
  \institution{German Research Center for Artificial Intelligence}
  \city{Kaiserslautern}
  \country{Germany}}
\email{hans_dieter.schotten@dfki.de}
\additionalaffiliation{%
  \institution{Department of Electrical and Computer Engineering, Technische Universität Kaiserslautern}
  \city{Kaiserslautern}
  \state{Germany}
}
\renewcommand{\shortauthors}{Daniel Reti, Karina Elzer, Daniel Fraunholz, Daniel Schneider, Hans Dieter Schotten}

\begin{abstract}
In the field of network security, with the ongoing arms race between attackers, seeking new vulnerabilities to bypass defense mechanisms and defenders reinforcing their prevention, detection and response strategies, the novel concept of cyber deception has emerged. Starting from the well-known example of honeypots, many other deception strategies have been developed such as honeytokens and moving target defense, all sharing the objective of creating uncertainty for attackers and increasing the chance for the attacker of making mistakes.
In this paper a methodology to evaluate the effectiveness of honeypots and moving target defense in a network is presented.
This methodology allows to quantitatively measure the effectiveness in a simulation environment, allowing to make recommendations on how many honeypots to deploy and on how quickly network addresses have to be mutated to effectively disrupt an attack in multiple network and attacker configurations. With this optimum, attacks can be detected and slowed down with a minimal resource and configuration overhead. 
With the provided methodology, the optimal number of honeypots to be deployed and the optimal network address mutation interval can be determined. 
Furthermore, this work provides guidance on how to optimally deploy and configure them with respect to the attacker model and several network parameters.
\end{abstract}

\keywords{network security, cyber deception, moving target defense, honeypot}

\maketitle

\section{Introduction}
In the past decades, many effective strategies for improving the security of Information Technology (IT) have come forward. Whether on the network level such as firewalls, Intrusion Detection Systems (IDS) and Intrusion Prevention Systems (IPS) or on the system level such as kernel hardening, antivirus and User Access Control. Nonetheless, many reports show that networks and systems are still being compromised despite the security measures in place. There are multiple reasons that are complex to mitigate, such as zero-day-exploits or malicious or naïve insiders. Thus, it is becoming clearer that concepts for defense-in-depth and resiliency need to be researched and ultimately deployed to aid the overall level of security. In parallel, the field of cyber deception is gaining relevance, as it addresses multiple problems in state of the art cyber defense. Cyber deception is seeking to distract adversaries from real targets, and present simulated targets instead, such as honeypots. This also addresses the traditional asymmetry between attackers and defenders, meaning that the defender has to be successful all of the time and the attacker is only required to be successful once.
In this paper, a network attack simulation is used to quantify the effectiveness of the presence of honeypots and Moving Target Defense (MTD), i.e. mutated network addresses, in networks of various sizes. The quantification is done by defining three attack strategies and for each strategy it is measured how likely it is for an attacker to successfully exploit a vulnerable host, when the different deception setting are in place.
The contribution of this paper is the following:
\begin{itemize}
  \item A methodology for optimizing honeypot and MTD deployment and effectiveness based on an enhanced version of a network attack simulation platform
  \item Introduction of various configuration options for the simulation to enable researcher and practitioners to tweak our methodology to their needs
  \item A quantitative study, comparing of the impact of various network, honeypot and MTD parameters based on the presented methodology 
\end{itemize}

The remainder of this work is structured as follows:
Section 2 gives an overview on the state of the art in cyber deception research. In Chapter 3 our Methodology is described, followed by the simulation environment in Chapter 4. In Chapter 5 the results are presented and a conclusion is given in Chapter 6.

\section{Cyber Deception}
The term \textit{cyber deception} describes a defense strategy of cyber security that is based on simulating fake targets and hiding real targets. While many cyber deception strategies exist, this section focuses on the popular techniques of honeypots and moving target defense, as those are used in this work.
An more extensive overview is presented by Fraunholz et al. \cite{fraunholz.2018}.

\subsection{Honeypots}
Honeypots are simulating hosts on a computer network which only serve as a decoy for attackers. Therefore any connection with a honeypot may be considered suspicious. A distinction can be made between low-, mid- and high-interaction honeypots\cite{fraunholz.2018}. While low interaction honeypots can reliably detect a host discovery scan and simple service interactions, a high interaction honeypot allows the attacker to take over and fully interact with the operating system. This allows researchers to learn about the behavior and capabilities of the attacker. A medium interaction honeypot is a compromise between high- and low interaction honeypot and typically allow interaction with services but not with the operating system. A distinction can also be made between production honeypots and research honeypots. While production honeypots are popular in corporate networks for attack detection and gaining insight on attacks, research honeypots can gather more attack specific data and involve higher maintenance costs. Many honeypot projects exist, differing in respect to which service they are emulating, how they are deployed and how much interaction is possible.
Well known honeypot projects include Kippo \cite{kippo}, Honeyd \cite{honeyd}, Glastopf \cite{Glastopf} and Conpot \cite{conpot}, which simulate open SSH ports, TCP ports, Web Applications and industrial control systems, respectively.

Another variation of honeypots, which do not simulate a host, but rather specific details, have been coined \textit{honeytoken} \cite{barros}. Such honeytoken are simulating a digital asset which looks interesting to the attacker. These can be for example decoy files, databases, credentials, or websites \cite{yuill2004honeyfiles} \cite{juels2013honeywords}. Honeytoken, like honeypots, make the reconnaissance phase, the earliest stage of an attack, difficult by creating uncertainty and increasing the chance of being detected.  
This uncertainty was proposed to be exploited, by proposing to make production systems appear like honeypots, namely fake honeypots \cite{rowe2007defending}.

\subsection{Moving Target Defense}
Similarly to the objectives of honeypots, MTD is aiming to create uncertainty for the attacker and waste their resources. To achieve these objectives, in MTD the attack surface is constantly shifting. This can take various shapes and is already established for Address Space Layout Randomization (ASLR) to better protect the memory from buffer overflow attacks or in radio networks channel hopping is used against jamming \cite{navda2007using}. Enabled through Software Defined Networking, constant mutation of network addresses, also referred to as random host mutation, has been proposed by different authors \cite{sharma.2018, sharma.2019, jafarian2012openflow, kampanakis2014sdn, fraunholz2018catch}. 
However, as with most security measures, the trade-off between security and performance needs to be taken into account \cite{kechao.2019,al-shaer.2013}. 
An extensive survey on MTD is presented in \cite{sengupta.2020}, highlighting the use of MTD against Advanced Persistent Threats and noting the extensive use of Artifical Intelligence and game theory in existing MTD approaches.

\section{Methodology}
In this section, the three attacker models are explained that have been defined for further examination. Next, the simulation environment used for the data collection is described. The environment includes a state of the art simulation tool as well as the changes which have been implemented to make Honeypots and MTD possible. The overall methodology to the used approach is described as well.

    \subsection{Attacker Model}
    \label{attackermodel}
        Three attacker agents with different assumptions on their behaviour have been defined and implemented. Each agent had a different tactic to attack the network. This tactic is implemented through fixed behaviour rules. The rules are separated into different phases. Parts of the decisions are chosen randomly. These choices allowed for a comparison between the agents.
        
        The actions an agent can chose from are the following:
        \begin{itemize}
        \item Subnet scan: Find out which addresses on the network are in use for host discovery
        \item Service scan: Find open TCP ports on a given host and derive an estimation on which services are present on that host.
        \item OS scan: Fingerprint the operating system and kernel version of the host.
        \item Vulnerability scan: Probe the target services for known vulnerability fingerprints
        \item Process scan: Used for privilege escalation, read which processes are running on a host.
        \item Exploit/Privilege Escalation: Chooses one of many predefined exploits
        \item Wiretapping: A post-exploitation step to capture unencrypted credentials from the network traffic.
        \end{itemize}
    
        The first agent is the 'careful' agent. This agent pursues a comprehensive horizontal scan, by first gaining an overview of the whole network and all attack possibilities before deciding on what action will be used for the attack. The 'careful' agents' actions are separated into the following phases:
        \begin{enumerate}
            \item The agent does a full set of all possible scans for all the existing hosts. This includes the subnet, service/port, vulnerability and Operating System (OS) scan.
            \item The agent chooses a random host. Then they pick the best exploit to gain user or root access. If the agent already has user access, it will do a privilege escalation instead.
            \item If the exploit was successful, the agent will perform a process scan if they obtained user rights. If the agent obtained root privileges, then the agent will do wiretapping.
            \item If Moving Target Defense is turned off or the agent does not detect it, they will repeat the steps from phase two until the end of the simulation. If Moving Target Defense is turned on, the agent will start with phase one.
        \end{enumerate}
        
        The second agent is called the 'standard' agent. For this agent, the design aimed toward a more realistic approach. The agent focuses on vertically scanning and attacking one host before looking at the next one. The phases for the standard agent are the following:
        \begin{enumerate}
            \item The agent performs a subnet scan to get the addresses of hosts connected to the network.
            \item The agent chooses a random host and only scans that host. They do a service, OS, vulnerability and process scan.
            \item Depending on the access level of the chosen host, the agent will perform an action from a list of exploits, privilege escalations or do wiretapping.
            \item If the action from phase three was successful, the agent goes back the phase two. Otherwise, the agent will continue in phase three until the host was successfully compromised the host or no exploits and privilege escalations are left. If Moving Target Defense is detected, the agent will start again from phase one.
        \end{enumerate}
        
        The last agent being defined is the 'aggressive' agent. This agent does not scan hosts. Instead, the agent performs attacks without knowing if they will be successful. This could be the behavior of a computer worm, which tries to exploit the same vulnerability on as many hosts as possible without the necessity to do a port scan. These are the phases for the 'aggressive' agent: 
        \begin{enumerate}
            \item The agent performs a subnet scan.
            \item The agent chooses a random action from the list of exploits and privilege escalations.
            \item The agent will perform the action on all hosts in a random order.
            \item If the attack was successful, the agent will do wiretapping, and then it will go back to phase two.
        \end{enumerate}

    \section{Simulation Environment}
        To gather the data for analysis, NASim was used. Honeypots and Moving Target Defense were added as features. In this chapter, the implementation of these additions is described, as well as the parameters chosen for the simulation runs. 
    
        \subsection{NASim}
        The NASim \cite{Schwartz.2020}, short for Network Attack Simulator, is a program to simulate network penetration testing. It is a python program that implements Open AI Gym \cite{Brockman16} to allow reinforcement learning. For this paper, this functionality has not been used. Instead, the capabilities of creating simulated computer networks and creating custom agents have been used. These agents have a clearer behaviour, which makes the comparison more comprehensibly.

        NASim allows the creation of a network with variable amounts of subnets, hosts and firewalls. It includes scanning of subnets, services (port scans), processes and operating systems (OS). Furthermore, it implements attacks through exploits and privilege escalations. For each of these actions, it is possible to assign costs.
        
        NASim defines a special type of host called a sensitive host. The agent's goal is to gain root access to all sensitive hosts. The simulation is considered successfully completed when the agent has exploited all sensitive hosts.
        
        The modifications were built on an already expanded version of NASim. In this previously extended version, the authors already implemented specific vulnerabilities, a vulnerability scan as well as the concept of credentials and wiretapping as an attack for obtaining credentials.
        
        \subsection{Implementation of Honeypots and Network Address Mutation}
            For this paper, the functionality of NASim was expanded to include Honeypots. Honeypots are based on the implementation of sensitive hosts. If an agent exploits one honeypot host with either user or root privileges, the goal failed and the agent has lost.
    
            Also, a second definition of the goal for the analysing was added. Instead of the agent needing to gain root privileges on all sensitive hosts, the second definition only needs one sensitive host to be successfully exploited for the agent to complete the goal.
            
            Additionally to the Honeypot feature, a  Moving Target Defense functionality was added. After a chosen time, hosts in the same subnet will randomly change their addresses. The time is defined by the actions taken and their associated costs. After this change, the agent does not know which hosts were already exploited.
    
            To make more addressed available for the mutation, 'empty' hosts were introduced. 'Empty' hosts are not detectable or attackable. Their only purpose is to introduce additional addresses in the subnets. Thus only serve as an unused address on the network, and the exiting hosts can shift to theses adresses. Therefore the address space is bigger and the simulation is more realistic.

    \section{Experimental Setup}
        For the test procedure, three different scenarios, based on different seed, were generated. It was decided which parameters of the simulation will be observed. Additionally, the termination conditions were set, defining when the agents would win and when they would lose.
    
        First, a modified version of the random scenario generator from NASim was used to generate the scenarios. The scenario generator allows for great customisation of the environment with slight randomisation and without the need to create the scenarios manually. 
        
        The most important parameters for this paper are the ones which were decided to be changed and observed. These are the Honeypots in the network and the time until address mutation of the MTD takes place. These two parameters and their comparison among themselves and with each other build the main focus of our research work presented. The objective defines whether the attacker needs to successfully exploit one or all sensitive hosts to win. Furthermore, the number of active normal hosts in the network, the agents as described in section \ref{attackermodel} and different random seeds that lead to different scenarios were identified. (Table \ref{tab:params})

        \begin{table}[t]
        \centering
        \begin{tabularx}{\linewidth}{|l|*{3}{>{\RaggedRight\arraybackslash}X|}}
            \hline
            Parameter & Description & Value \\
            \hline
            \hline
            num\_honeypots\_options & Number of Honeypots & 0, 2, 4, 6, 9, 10 \\
            \hline
            movement\_time\_options & Movement Time & None, 25, 50, 75, 100 \\
            \hline
            num\_hosts\_options & Number of normal hosts & 10, 50 \\
            \hline
            one\_goal\_options & Objective & True, False \\
            \hline
            seed\_options & Random Seed & 1234, 42, 24121997 \\
            \hline
            agents & Attack behaviour of agents & careful, standard, aggressive \\
            \hline
        \end{tabularx}
        \caption{All Possible Varied Parameter Values}
        \label{tab:params}
        \end{table}
        
        \begin{table}[t]
        \centering
        \begin{tabularx}{\linewidth}{|l|*{3}{>{\RaggedRight\arraybackslash}X|}}
            \hline
            Parameter & Description & Value \\
            \hline
            \hline
            num\_sensitive & Number of sensitive hosts & 3 \\
            \hline
        	num\_services & Number of services & 10 \\
        	\hline
        	num\_os & Number of OS & 1 \\
        	\hline
        	num\_processes & Number of processes & 10 \\
        	\hline
        	num\_exploits & Number of exploits & 10 \\
        	\hline
        	num\_privescs & Number of privilege escalations & 10 \\
        	\hline
        	num\_vulns & Number of vulnerabilities & 10 \\
        	\hline
        	num\_creds & Number of credentials & None \\ 
        	\hline
        	r\_sensitive & Reward for sensitive hosts & 1000 \\
        	\hline
        	r\_honeypot & Reward for honeypots & -1000 \\
        	\hline
        	*\_cost & Action costs & 1 \\
        	\hline
        	exploit\_probs & Success probability for exploits & 1.0 \\
        	\hline
        	privesc\_probs & Success probability for privilege escalations & 1.0 \\
        	\hline
        	uniform & Uniform distribution of hosts & True \\
        	\hline
        	base\_host\_value & Reward for normal host & 1 \\
        	\hline
        	host\_discovery\_value & Reward for host discovery & 1 \\
        	\hline
        	step\_limit & Maximum number of actions permitted & 3000 \\
        	\hline
        	addresses & Number of network addresses & 256 \\
        	\hline
        	subnets & Number of subnets & 2 \\
        	\hline
        \end{tabularx}
        \caption{All Fixed Parameter Values}
        \label{tab:fixedparams}
        \end{table}
        The rest of the parameters were fixed throughout all simulations. The network included 256 addresses and two subnets, one only for the attacker. Since the focus was not layed on the firewall, no restrictions were included. The costs of each action was set to one. A distinction was made between the reward for the sensitive, honeypot and normal hosts. Three hosts were defined as sensitive hosts and for each scenario ten possible services, processes, exploits, privilege escalations and vulnerabilities were given. The number of Operating Systems was set to one and credentials have not been considered. (Table \ref{tab:fixedparams})

        The simulations can terminate in three different ways:
        \begin{itemize}
            \item The agent wins, which means the agent successfully exploited all necessary sensitive hosts. This is dependent on the 'one\_goal' parameter if the agent needs to exploit only one sensitive host or all three.
            \item The agent loses. This means the agent exploited a honeypot.
            \item The agent takes too long, which is defined by the 'step\_limit' parameter.
        \end{itemize}
        
        Finally, after choosing all parameters, we ran the simulations. Each combination of parameters was run 100 times to compensate for randomness in the scenarios and agent behaviours. 

\section{Results} %
\label{sec:results}
As described in the previous chapters, the simulation was done by the three types of attackers in randomly generated network scenarios. Hereby different settings were tested, with a varying number of hosts, honeypots and network address mutation frequencies. In the following, the results from our experiments are presented from three different perspectives, starting with a comparison of the different attacker types, followed by the impact of honeypots and lastly the impact of moving target defense is considered. It is important to state the results reflect an assessment of honeypots and MTD considering an example configuration, and thus can be adopted to any network specifics using the proposed methodology to suit the needs of researchers and practitioners.

\subsection{Agent Performance} 
\begin{figure}[ht]
    \centering
    \includegraphics[width=0.5\textwidth]{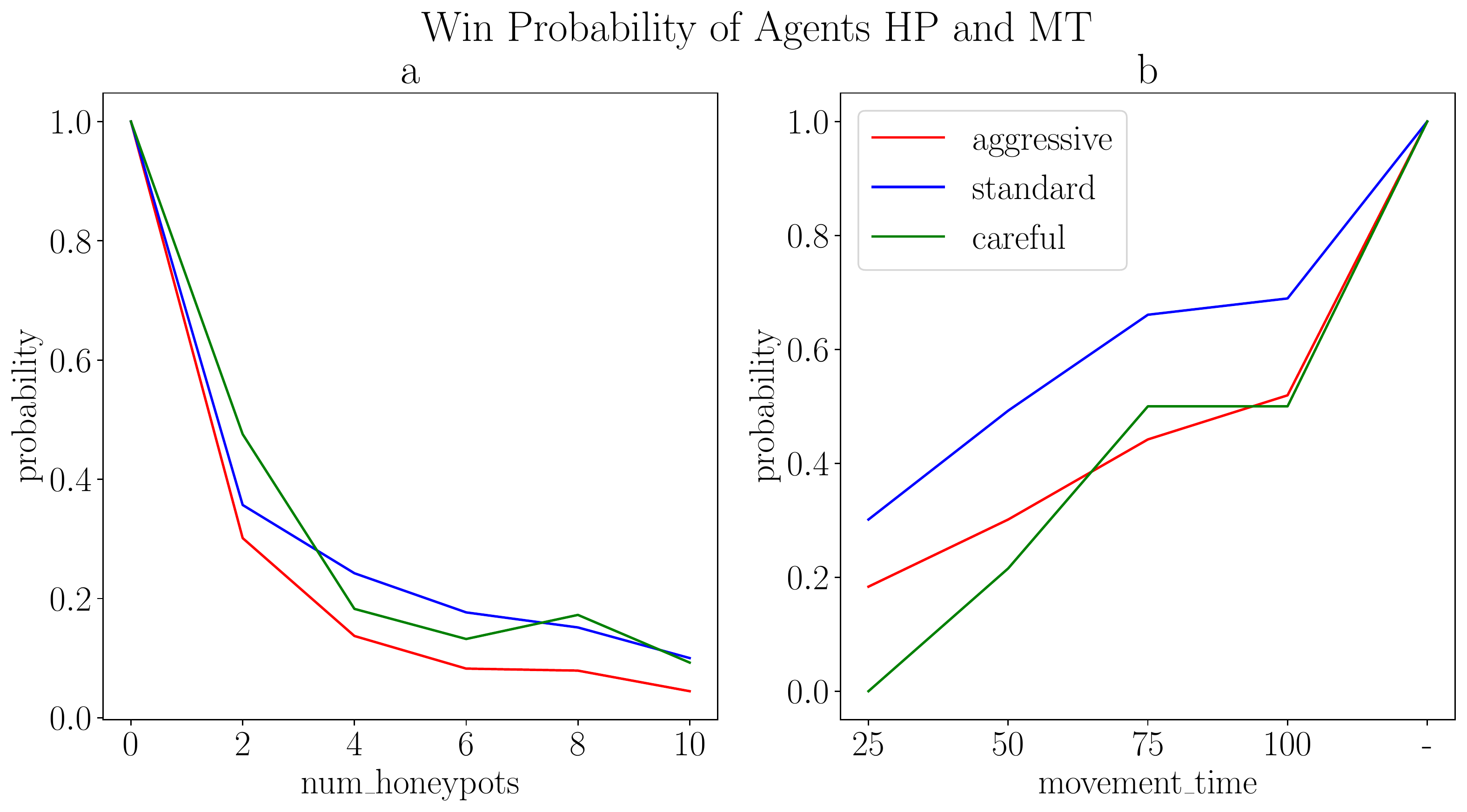}
    \caption{Comparison of the winning probability for the three agents for a) an increasing number of honeypots and b) an increasing IP changing interval.}
    \label{fig:mt_barcharta}
\end{figure}

The three different attacker types implemented in the simulation were a careful agent, who does a full scan of the network before it runs attacks, a standard agent who will scan single hosts and run an attack when a vulnerability is identified and the aggressive agent who will try a single attack on all hosts in the network. An agent will lose, when a honeypot is exploited or the step limit, as presented in Table \ref{tab:fixedparams}, is exceeded. Therefore an increasing amount of honeypots in the network will decrease the chances of all three agents to win. %
But also the careful agent is highly affected as it does a horizontal scan and might identify a honeypot as a potential targets. As can be seen in Figure \ref{fig:mt_barcharta}a, the winning probability of the standard agent is falling earlier with an increasing number of honeypots, while careful agent's winning probability dropping below the standard agent at around 4 honeypots and very close to it with 10 and above. In respect to MTD, as shown in \ref{fig:mt_barcharta}b, moving target defense, with a lower movement interval of 25, the careful agent's winning probability is 0, as the IP mutation is happening before the agent chooses to attack. The aggressive agent is performing better for movement intervals below 60, and close for higher intervals. The standard agents has throughout the highest winning chance,  which is due to relatively quick actions on a randomly picked target.

\begin{figure}[ht]
    \centering
    \includegraphics[width=0.5\textwidth]{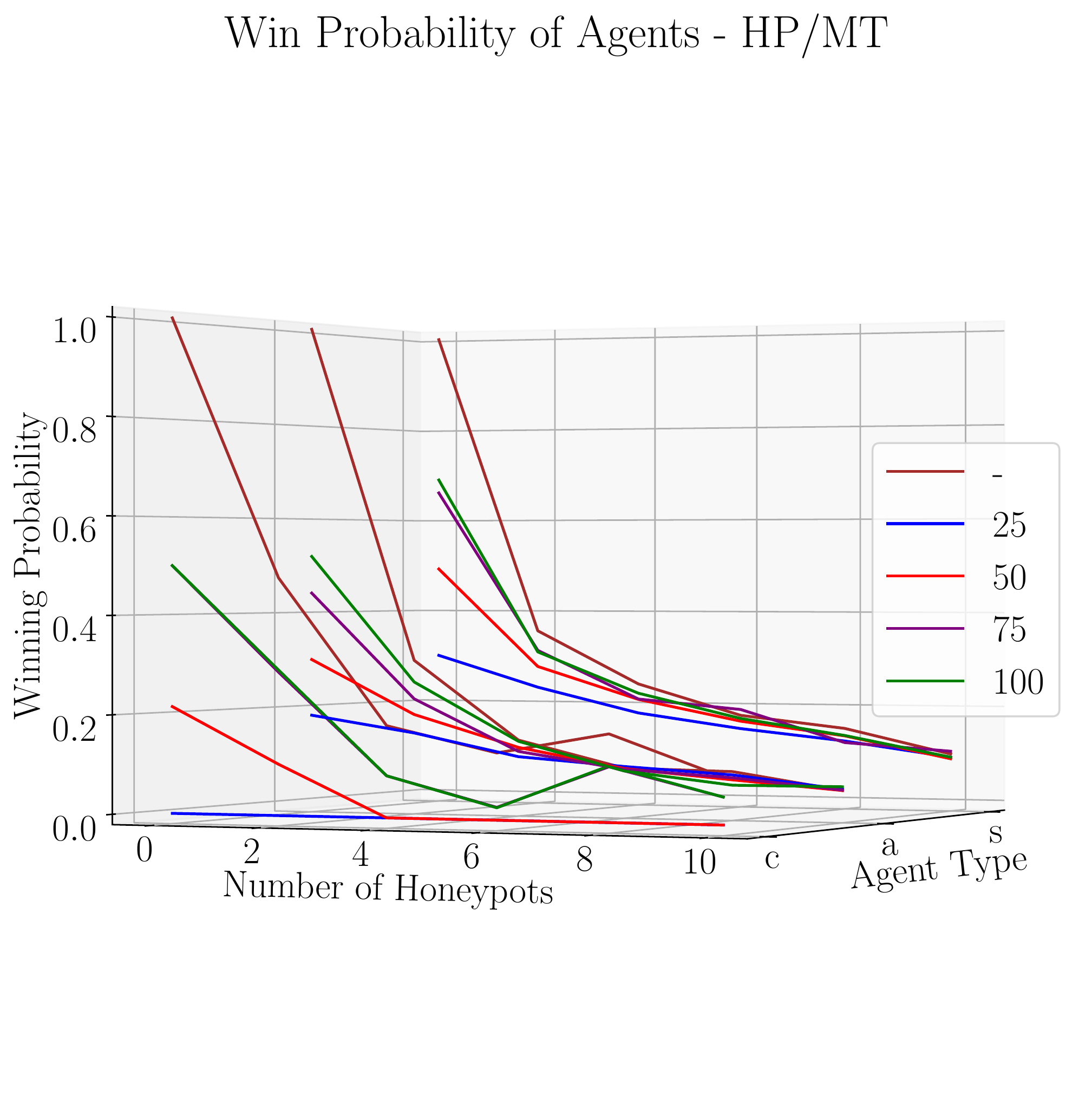}
    \caption{Winning probability for different IP changing times of the standard agent (s), aggressive agent (a) and careful agent (c) over an increasing number of honeypots (z-axis)}
    \label{fig:agents_comparison_3d}
\end{figure}

In Figure \ref{fig:agents_comparison_3d}, the agents are compared with honeypots and moving target defense enabled simultaneously. All three agents have a decreasing winning probability with lower mutation intervals and with an increasing amount of honeypots. It can be seen, that the standard agent is performing slightly better on average, with a winning probability of 44\%, compared to 26\% with the careful and 30\% with the aggressive agent. %

\begin{figure*}[ht]
    \centering
\begin{subfigure}[t]{.48\textwidth}
    \centering
    \includegraphics[width=.95\linewidth]{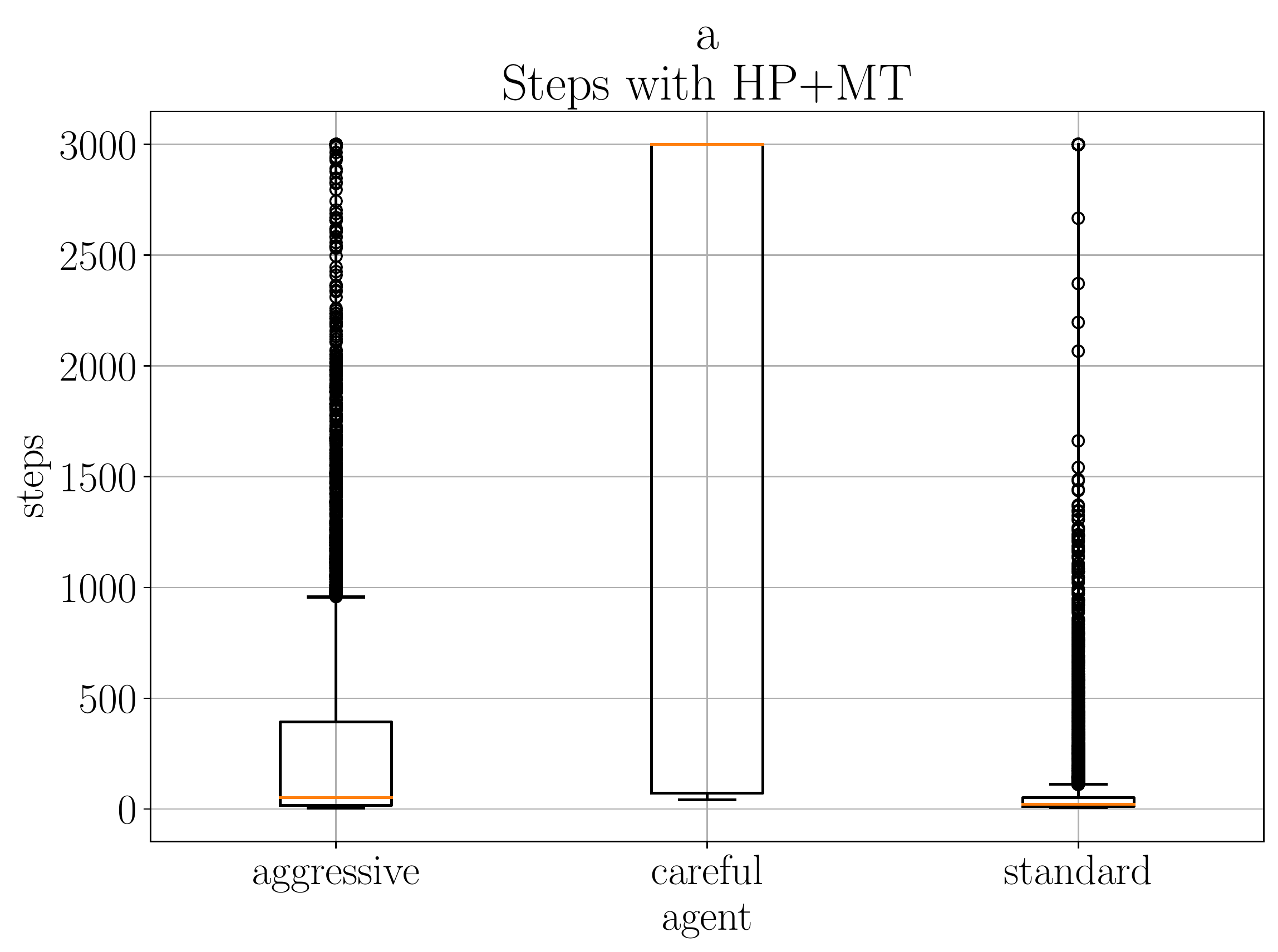}
    \label{fig:boxplot_w_hp_mtd}
\end{subfigure}%
\begin{subfigure}[t]{.48\textwidth}
    \centering
    \includegraphics[width=.95\linewidth]{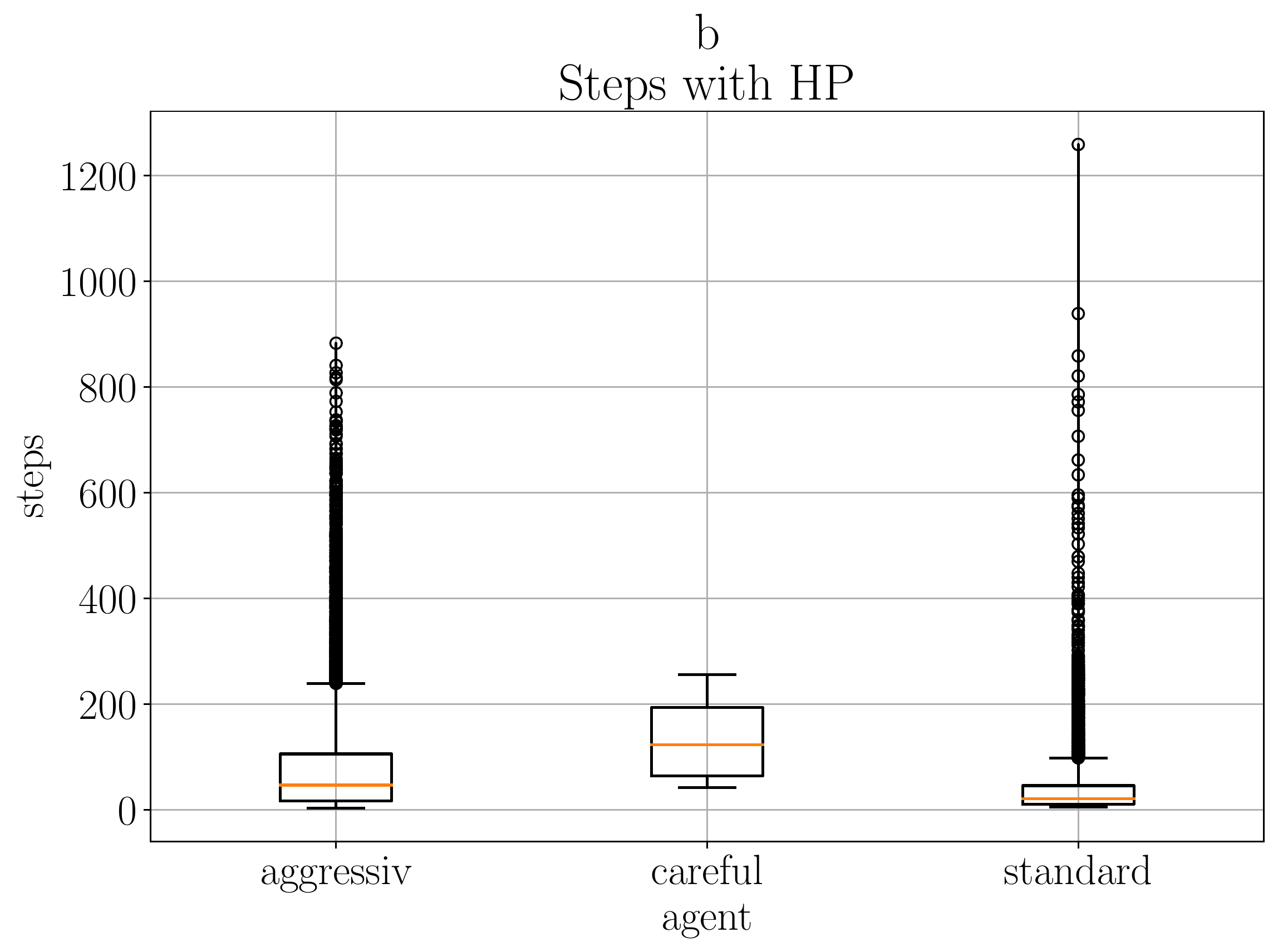}
    \label{fig:boxplot_wo_hp_mtd}
\end{subfigure}
\begin{subfigure}{.48\textwidth}
    \centering
    \includegraphics[width=.95\linewidth]{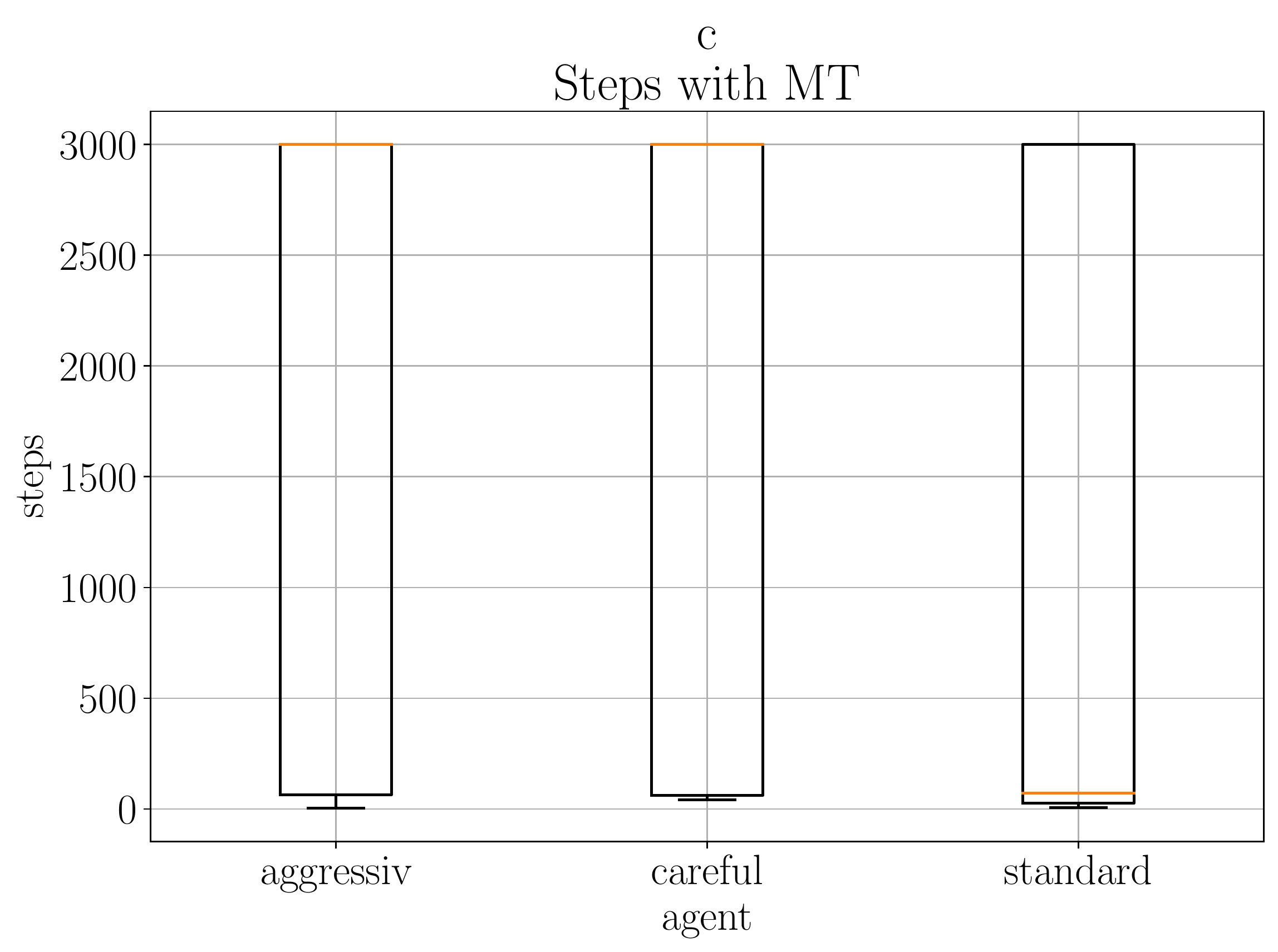}
    \label{fig:boxplot_w_hp}
\end{subfigure}
\begin{subfigure}{.48\textwidth}
    \centering
    \includegraphics[width=.95\linewidth]{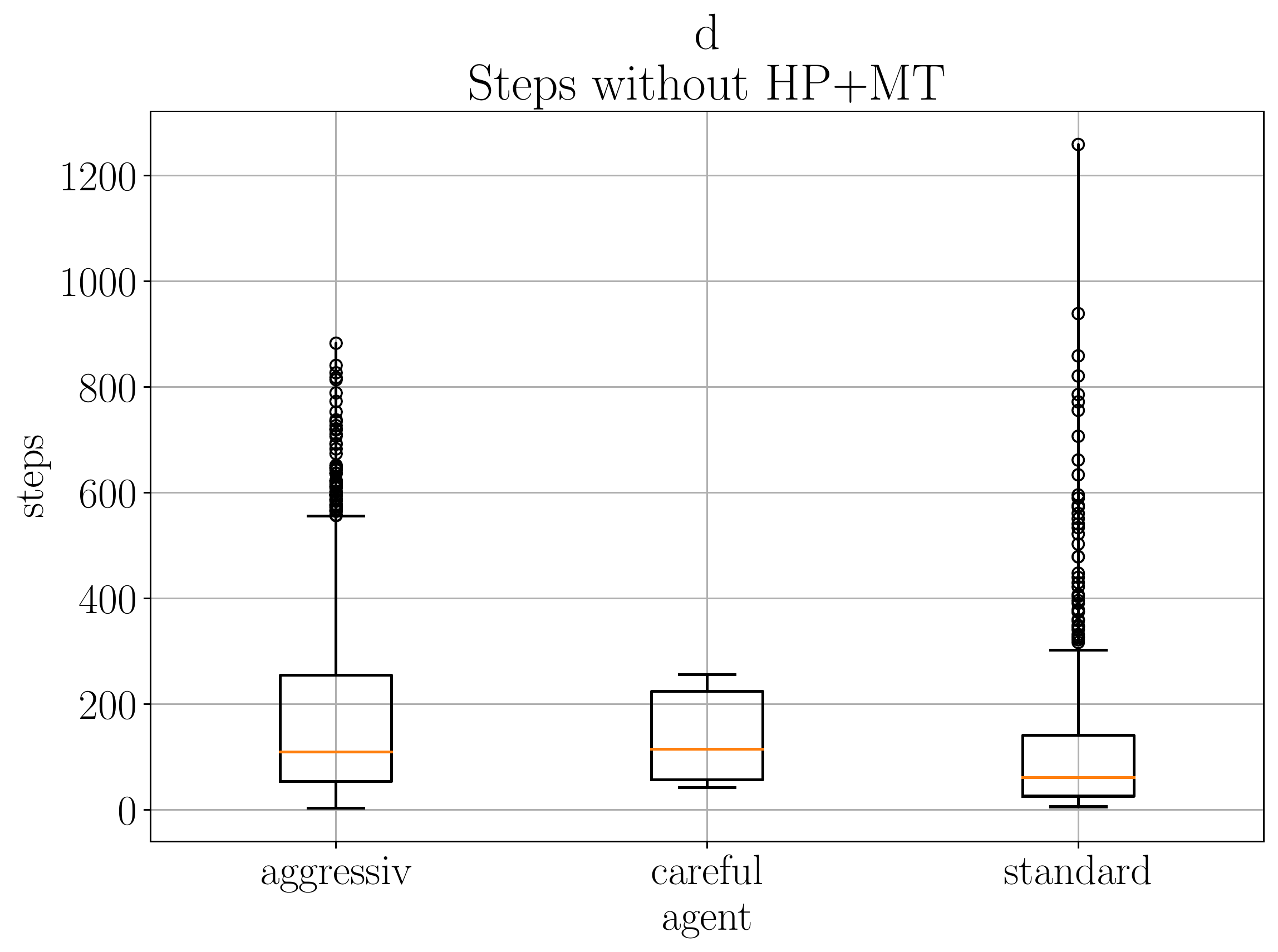}
    \label{fig:boxplot_w_mtd}
\end{subfigure}

    \caption{Comparison of steps needed for the three agents, a) MTD and Honeypots, b) Honeypots only, c) MTD only, d) no deception measures.}  \label{fig:agents_comparison_boxplot}  

\end{figure*}

Focusing on the number of steps needed by each agent, it can be seen in Figure \ref{fig:agents_comparison_boxplot}, that when honeypots and MTD are in place, that the aggressive agent has its maximum below 1000 steps, the careful agent performs poorly with its median at 3000 steps, which means that it doesn't succeed most of the time, and the standard agent, which performs the best with a median and a maximum below 100 steps. For reference, when neither honeypots nor MTD is enabled, all three agents have a median close to 100 steps. When only honeypots are enabled, the number of steps drops, as an interaction with the honeypot end the round. When only MTD is enabled, the median of the aggressive and the careful agent rise to the maximum, while the standard agent has a low median, while its third quartile is at the maximum value of 3000 steps.

\subsection{Honeypots}
\begin{figure}[ht]
    \centering
    \includegraphics[width=0.475\textwidth]{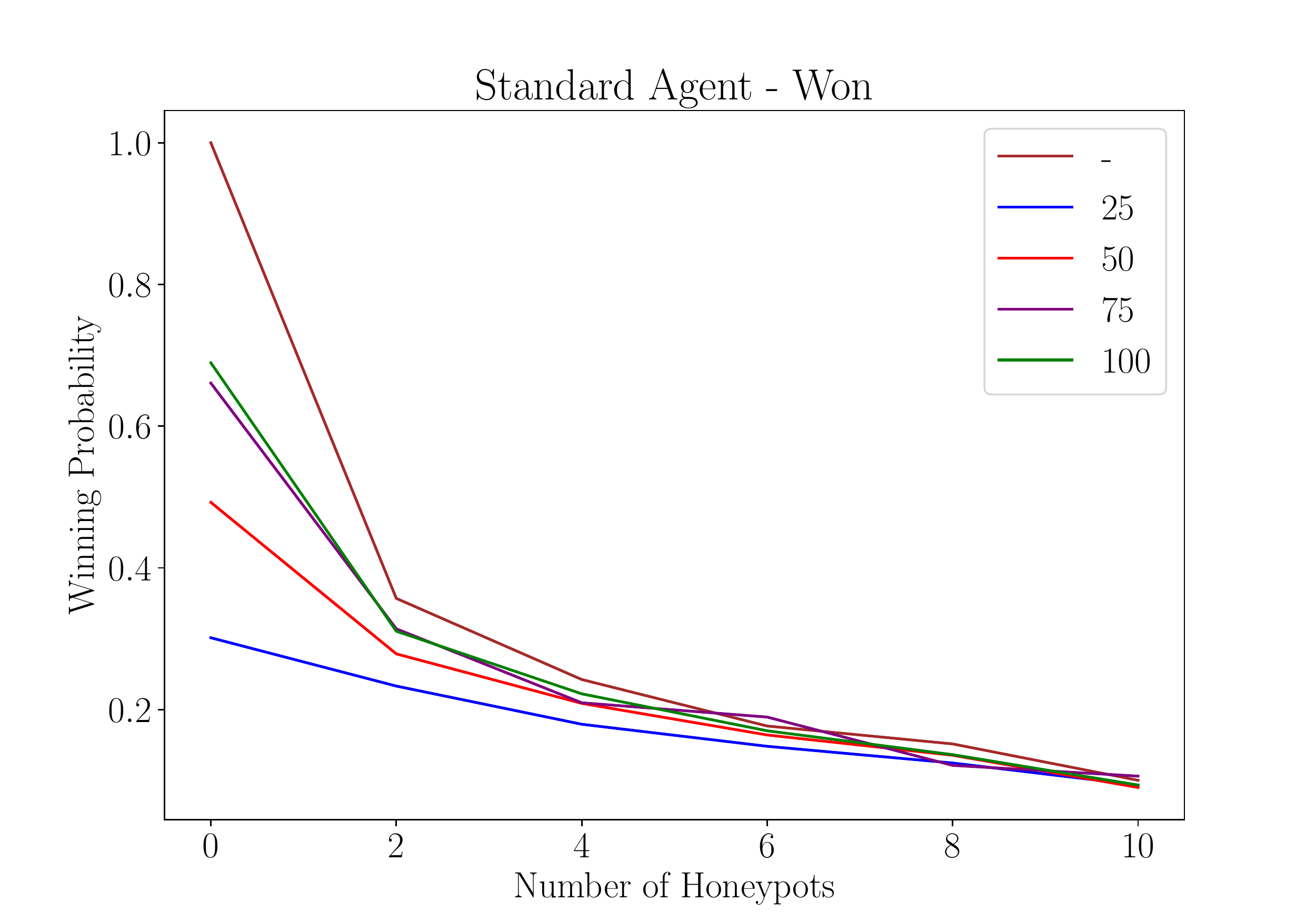}
    \caption{Comparison of the probabilities for the standard agent to win for an increasing number of honeypots. The colors represent different movement times}
    \label{fig:honeypot_winprob}
\end{figure}

\begin{figure}[ht]
    \centering
\begin{subfigure}{0.49\textwidth}
    \centering
    \includegraphics[width=\textwidth]{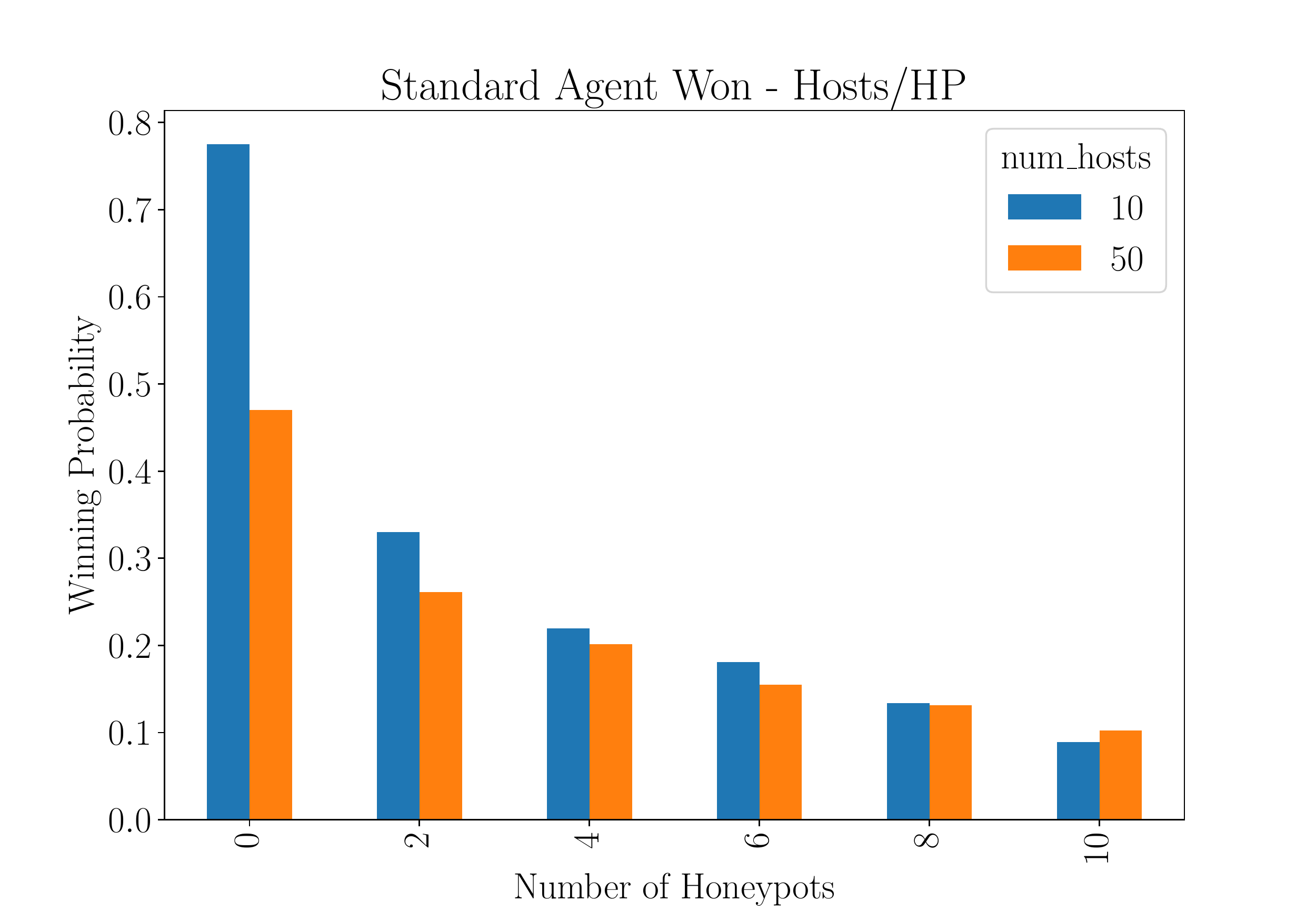}
    \caption{Comparison of the probabilities for the standard agent to win for an increasing number of honeypots in networks  of 10 and of 50 hosts.}
    \label{fig:hp_barchart}
\end{subfigure}

\begin{subfigure}{.5\textwidth}
    \centering
    \includegraphics[width=\textwidth]{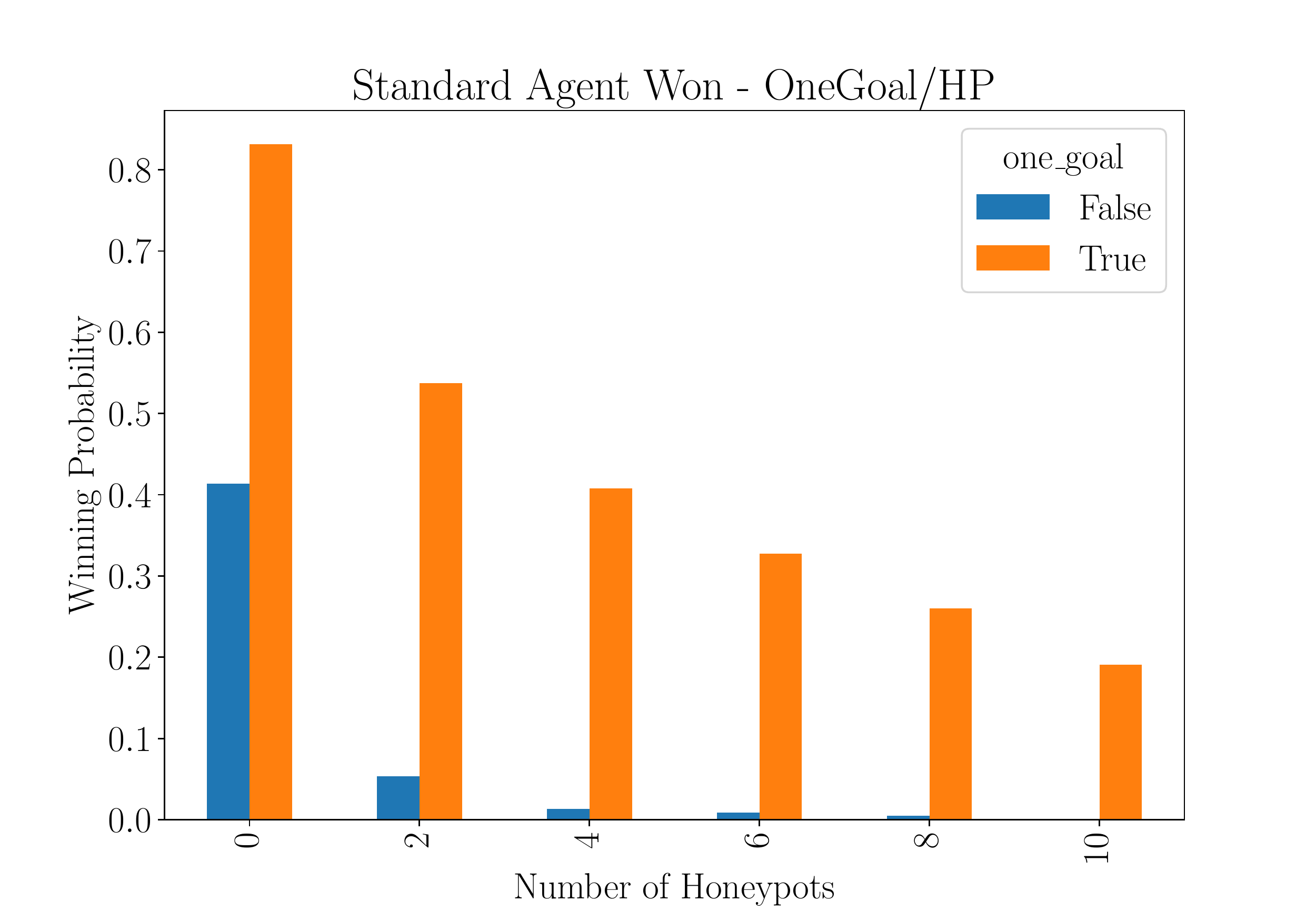}
    \caption{Comparison between the probabilities of the standard agent to compromise one host and to compromise all hosts for an increasing number of honeypots, respectively. }
    \label{fig:hp_onegoal}
\end{subfigure}
    \caption{Comparison between the winning probabilities of the standard agent for different configurations.}
    \label{fig:barcharts2}  
\end{figure}

In the previous section, the impact of honeypots on the different attack strategies was discussed. In this section various honeypot configurations are compared for a single attack strategy. In Figure \ref{fig:honeypot_winprob}, the impact of 0 to 10 honeypots is compared for different mutation frequencies of 25, 50, 75 and 100 time steps. With a number of 2 honeypots, the winning probability of the agent is already decreased from 100\% to below 40\% for all mutation intervals. Employing 8 honeypots further decreases the probability to below 20\% and 10 honeypots to below 10\%. An observation can be made, that with an increasing number of honeypots the mutation interval is having less influence on the winning chance. This also implies that less perturbations are needed the more honeypots are in place. To see the effect of honeypots without host mutation, \ref{fig:hp_barchart} shows the probability to find and exploit all vulnerable hosts, for 0 to 10 honeypots in a network of 10 and 50 hosts respectively. This show that even in a small network of 10 hosts, the employment of 2 honeypots is enough to decrease the winning probability from 77\% down to 34\%. As depicted in Figure \ref{fig:hp_onegoal}, a comparison was made between the chance of exploiting all vulnerable hosts and exploiting at least one. While for compromising only one sensitive host the probability is 82\% compromising all three hosts is only  41\%. For two honeypots the chance for compromising one host decreases to 55\% and for three host decreases to 5\%. 

\subsection{Moving Target Defense}
\begin{figure*}[hbtp]
    \centering

\begin{subfigure}[b]{.475\linewidth}
    \centering
    \includegraphics[width=\textwidth]{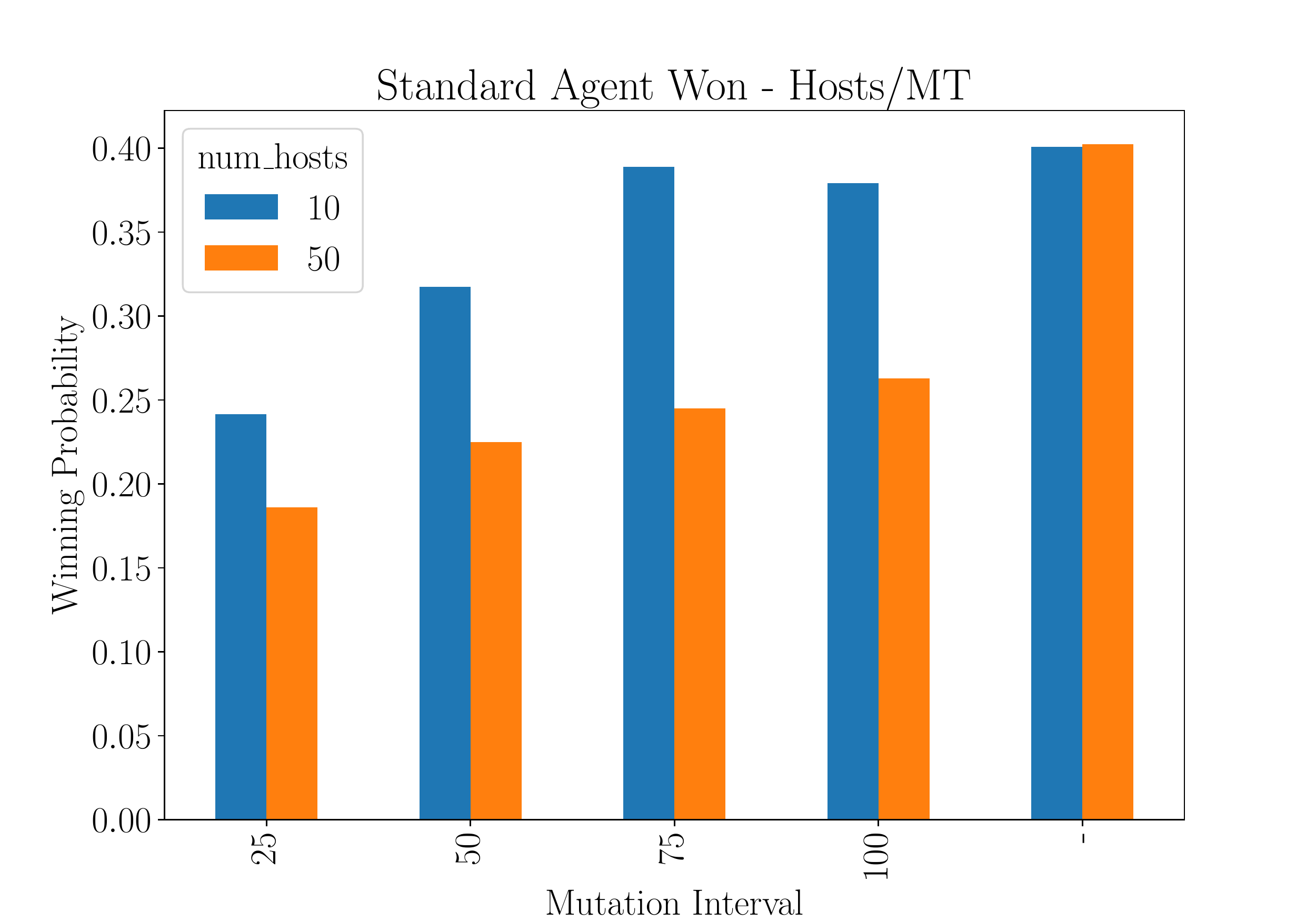}
    \caption{Comparison of the probabilities for the standard agent to win for an increasing IP changing interval in networks of 10 and of 50 hosts.}
    \label{fig:mt_barchartr}
\end{subfigure}\hfill%
\begin{subfigure}[b]{.475\linewidth}
    \centering
    \includegraphics[width=\textwidth]{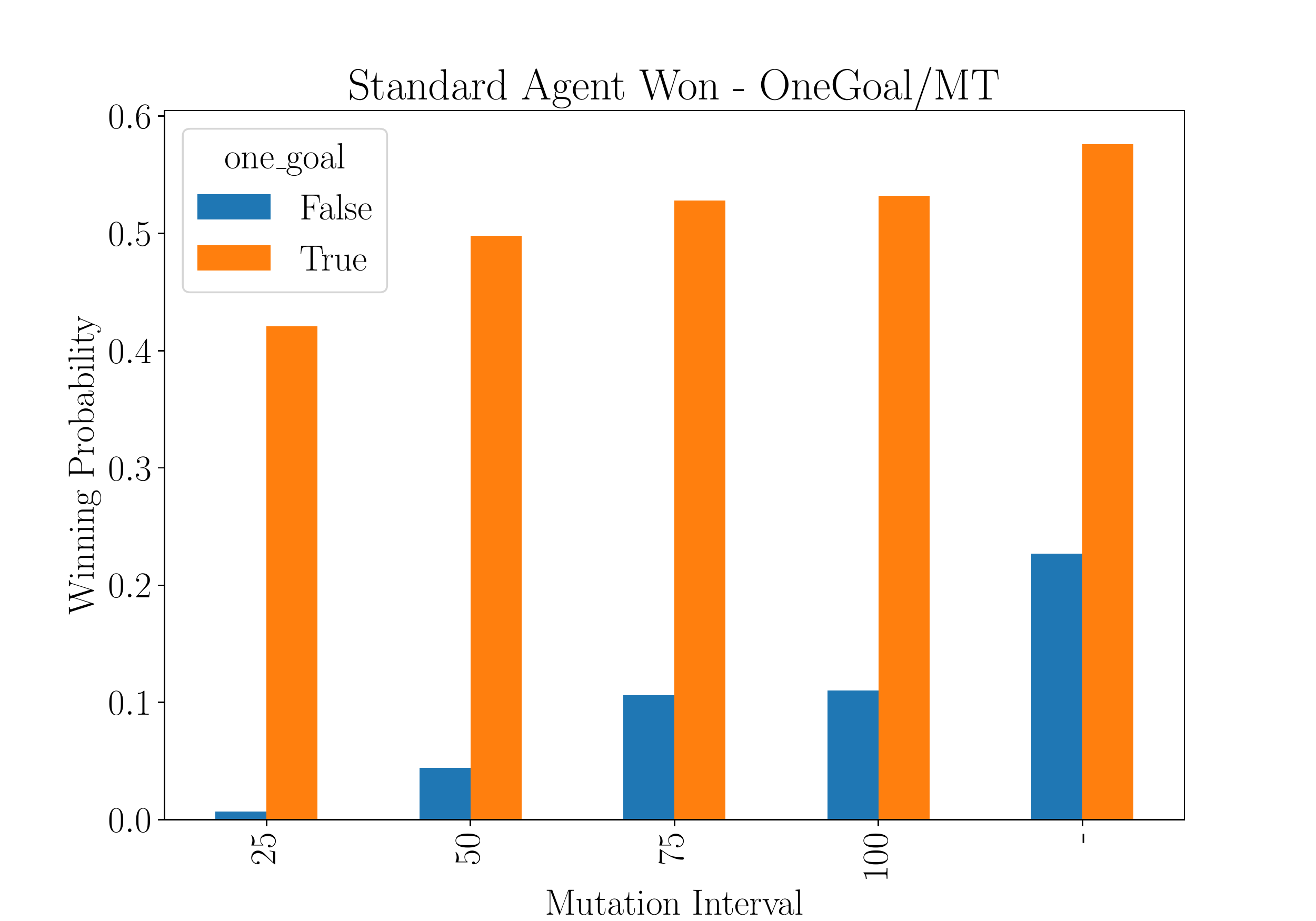}
    \caption{Comparison between the probabilities of the standard agent to compromise one host and to compromise all hosts for an increasing IP changing interval.}
    \label{fig:mt_onegoal}
    \end{subfigure}
    \caption{Comparison between the winning probabilities of the standard agent for different configurations.}
    \label{fig:barcharts}  
\end{figure*}

For the MTD, the evaluation was made by varying the mutation intervals, as can be seen in Figure \ref{fig:mt_barchartr}, for network sizes of 10 and 50 hosts, respectively. The results show that employing moving target defense every 25 time steps for a network of 10 hosts has the highest chance of winning with 78\%, while the a larger network of 50 host increases difficulty to find the vulnerable target and decrease the winning chance to 48\%. Increasing the movement time to 50 increases the winning probability to 33\% and 23\% respectively. For 10 hosts, the mutation interval of 75 is already very close to the reference of having no MTD in place. For larger networks of 50 hosts, the increase of movement intervals does slightly increase the chance of winning, with the highest movement interval still being 33\% lower than the reference of no MTD.

For only one goal, as depicted in \ref{fig:mt_onegoal}, the MTD does not decrease the winning chance below 40\% for any interval, as shown in Figure \ref{fig:mt_onegoal}. Meanwhile compromising all three sensitive hosts is below 1\% for the short mutation interval of 25 and at 10\% for the longer mutation interval of 100 time steps.

\section{Related Work}
While much research has been conducted on designing specific types of honeypots, decoys or moving target defense strategies, and the implications on attack detection and slow-down have been discussed, there is no standard strategy for the evaluation of such techniques.

Thus, the evaluation of deception technologies has been attempted in different forms. Many approaches are based on the formalization of the attacker and the defender using game theory \cite{pibil2012game}. Garg et al. created a framework  based on extensive games of imperfect information to determine strategies of the attacker and the honeynet system \cite{garg2007deception}. Quang Duy et al. modeled the attacker and defender as a Bayesian game of incomplete information, where both the attacker and the defender have deception capabilities, to determine the strategies of both defending on the attack frequency \cite{QuangDuy2016}. Hereby the attacker can obfuscate attacks and the defender can incorporate honeypots.

Other works rely on experiments with human subjects acting as attackers, such as Han et al., who evaluated web based decoys resulting in a detection rate of 67\% of the 150 participants with a low rate of false positives \cite{Han2017} or Ferguson-Walter et al., who used a psychological approach to evaluate deception by measuring the cognitive load on human penetration testers while interacting with cyber deception \cite{Ferguson2021}. Bensalem et al. considered deception as a defense against masquerade attacks, in which attackers pose as legitimate users, e.g. in identity theft \cite{bensalem.2011}. Their method of deception consisted of monitored honeyfiles acting as a IDS. The evaluation was done by conducting scenario-based user studies, in which computer science students were tasked to play the role of a disgruntled employee, leveraging physical access to a co-workers workstation for illicit financial gain. A special focus of the study were the properties of the honeyfiles. Heckman conducted a cyber-wargame, in which deception was employed along with a certificate-based defense strategy \cite{heckman.2013}. In this setup, a team of attackers tried to gain access to a command and control system defended by two defending teams, each of which employed one of the mentioned defense strategies, with the certificate-based strategy being the first line of defense, and the deception strategy only being activated once the attackers breached the system, triggering a switch of the target system to a backup instance. The results of this experiment showed that while certificate-based defense could not deny the attacking team access to the target system, the deception-based defense succeeded in leading the attackers to waste their time and resources on a decoy system.
Shabtai et al. studied insider data misuse by adding automatically generated honeytokens to collections of sensitive data, such as loan applications \cite{shabtai.2016}. In an experiment with student participants it was studied how well honeytokens pass for authentic data, and how well adversarial behavior can be detected by them. Interestingly, while honeytokens could indeed be shown to serve as efficient sensors for malicious activity, the knowledge of honeytokens being deployed did not influence the participants' behavior, although the authors do conclude that this might be due to a lack of real-life consequences in the experimental setting. 

Brewer et al. studied the defense of websites against automated bot attack by implementing a deception service that actively modifies a website's HTML code to include decoy links, which allow for the detection of bot activity. This service was evaluated by deploying three instances of websites protected by the service and then respectively attacking each website with a different type of bot, modeled after such bots as commonly found in the wild. The websites were protected, i.e. not publicly available, thus human volunteers were invited to visit the websites and generate non-malicious traffic, as to allow a distinction between malicious bot activity and legitimate user activity \cite{brewer.2010}.

In a simulation based approach, Ferguson-Walter et al. used the network attack simulator CyberBattleSim, developed by Microsoft, to evaluate the effectiveness of deception against a reinforcement learning (Rl) based penetration testing agent \cite{walter2021incorporating}. Hereby the focus lies on incorporating decoys in the simulation instead of honeypots, and measuring the effect of the number of the applied deceptive elements on the attack.

While the simulation environment used in this work is originally also intended for RL based agents, it was chosen to define the attackers with strict strategies to have more comparable and explainable results. The evaluation and recommendation of deception strategies is still unsolved. A simulation methodology combining honeypots and moving target defense, as described in this paper, has not been presented before to the author best knowledge.

\section{Conclusion}
In our work, a simulation based methodology was presented for evaluating the required configuration of honeypots and MTD on a network to achieve certain intrusion detection rates for defined attacker agents, depending on different network parameters that need to be provided, such as network size and an assumption about the amount of sensitive hosts, and the attacker type. Three different attacker types have been defined and considered and can be chosen for the simulation. The simulation, which was implemented based on Python and the NASim library, has shown to provide valuable insight on the impact of honeypots and MTD in various attack scenarios. 
The results can be utilized to determine the optimal trade-off between the number of honeypots or size of mutation intervals in a network and the resulting intrusion detection and prevention capabilities and on the other hand the increased resources required to configure, deploy and maintain the honeypot and MTD implementations. Also a lower number of honeypots saves IPv4 address space which might be reserved or used for other purposes. 
Although the simulation is based on abstraction and many assumptions have to be taken, the result is of value if these assumptions can be sufficiently justified for the real world scenarios of practitioners. Additionally,
researchers may use the proposed simulation methodology or implementation to evaluate their deception methods and tools in dynamic environments.
The results of the case studies provided in Section \ref{sec:results} of this paper are only valid for the specific simulation parameters and can not be seen as general metrics, since the parameters in the simulation, such as limiting an attack to 3000 steps, may not accurately describe any real world scenario. They should be rather interpreted in relation within the simulation setting to measure the impact of the attacker types, network size, mutation intervals and honeypots.

\begin{acks}
This research was supported by the German Federal Ministry of Education and Research (BMBF) through the Open6GHub project (Grant 16KISK003K).
\end{acks}

\bibliographystyle{ACM-Reference-Format}
\balance
\bibliography{literature}
  
\end{document}